\documentstyle[nato,epsf]{crckapb}

\newcommand{\be}{\begin{equation}}
\newcommand{\ee}{\end{equation}}
\newcommand{\bdm}{\begin{displaymath}}
\newcommand{\edm}{\end{displaymath}}
\newcommand{\<}{\langle}
\renewcommand{\>}{\rangle}

\def\ham{{\bf \rm H}}

\def\mbham{{\cal H}}
\def\bmat{{\bf \rm B}}
\def\cmat{{\bf \rm C}}

\def\su3{$SU(3)$}

\begin{opening}
\title{Topology and Chiral Symmetry in QCD with Overlap Fermions
\vskip-3.7cm\hfill \vbox{\hbox{\rm FSU-SCRI-99C-72} 
\hbox{\rm JLAB-THY-99-41}}\vskip2.3cm
}

\author{Robert G. Edwards}
\institute{
Jefferson Lab,
12000 Jefferson Avenue, \\
MS 12H2,
Newport News, VA 23606, USA}
\author{Urs M. Heller}
\institute{
SCRI, Florida State University, \\
Tallahassee, FL 32306-4130, USA}
\author{Joe Kiskis}
\institute{
Dept. of Physics, University of California, \\
Davis, CA 95616}
\author{Rajamani Narayanan}
\institute{
American Physical Society, \\
One Research Road,
Ridge, NY 11961, USA}

\end{opening}

\runningtitle{Topology and Chiral Symmetry $\dots$}

\begin{document}

\begin{abstract}
We briefly review\footnote{Talk given by Urs M. Heller at the workshop
``Lattice fermions and structure of the vacuum'', October 5--9, 1999,xi
 Dubna, Russia.}
the overlap formalism for chiral gauge theories, the
overlap Dirac operator for massless fermions and its connection to
domain wall fermions. We describe properties of the overlap Dirac operator,
and methods to implement it numerically. Finally, we give some examples
of quenched calculations of chiral symmetry breaking and topology with
overlap fermions.
\end{abstract}

\section{Overlap formula for the chiral determinant}

In the overlap formalism~\cite{overlap}, the chiral determinant is
obtained by embedding the Weyl fermion inside a Dirac fermion through
a many-body problem. Let $\mbham^\pm$ be two many-body Hamiltonians
\begin{eqnarray}
\mbham^\pm &=& - \pmatrix{ a^\dagger_1 & a^\dagger_2 }
 \ham^\pm \pmatrix{ a_1 \cr a_2 } \nonumber \\
\ham^- = \gamma_5 = \pmatrix{ 1 & 0 \cr 0 & -1 }; && \!\!\!
\ham^+ = \gamma_5 \left( \gamma_\mu D_\mu - m \right)
 = \pmatrix{ -m & \cmat(A) \cr \cmat^\dagger(A) & m } .
\end{eqnarray}

Let $|0\pm\rangle$ be the ground states of $\mbham^\pm$
obtained by filling all the positive energy states of $\ham^\pm$.
Then
\be
\det \cmat(A) \Leftrightarrow \<0-|0+\>_A
\ee

``Proof'': $|0-\rangle$ is obtained by filling all the positive energy states
of $\gamma_5$, and $|0+\rangle$ by filling all the positive energy states
of $\ham^+$. They are of the form:
\be
|0-\rangle ~:~ \pmatrix{ 1 \cr 0 }~, ~~~{\rm and} ~~~
|0+\rangle ~:~ \frac{1}{N_k} \pmatrix{ \cmat u_k \cr
\left( \sqrt{\mu_k^2 + m ^2} +m \right) u_k } ~,
\label{eq:0-+}
\ee
where $\cmat^\dagger\cmat u_k = \mu_k^2 u_k$ and
$N_k$ is the normalization.

The overlap formula is formal and needs to be regulated.
It is valid only in the limit $m\rightarrow\infty$ and one should think
of $m$ as a pre-regulator. The formula is strictly valid only for ratios of determinants
since there is a gauge field independent normalization in the formula.

$\ham^+$ need not have an equal number of positive and negative energy states
and this happens for topologically non-trivial
gauge fields, where the difference between the number of negative and positive
energy states of $\ham^+$ is $2Q$. Then $\det C(A)= 0$!
Furthermore,
\be
\<0-|a^\dagger_{i_1} \dots a^\dagger_{i_Q} |0+\> \qquad {\rm or} \qquad
 \<0-|a_{i_1}\dots a_{i_{|Q|}} |0+\>
\ee
will be non-zero, for $Q>0$ or $Q<0$, respectively, resulting in fermion number violation.
Potenital anomalies reside in the phase of $|0+\>$.
We will be concerned only with vector gauge theories, where only
$|\<0-|0+\>|^2$ enters and the anomaly is trivially cancelled in this case.

\section{Lattice regularization}

On the lattice $\ham^- \to \ham_L^- = \gamma_5$ remains unchanged,
while
\be
\ham^+ \to \ham_L^+ \equiv \ham_w(m) = \gamma_5 D_w(-m) = \pmatrix{
 \bmat(U)-m  & \cmat(U) \cr \cmat^\dagger(U) &  -\bmat(U) + m }
\ee
where $\cmat(U)$ is the naive lattice discretization of $\cmat(A)$
and $\bmat(U)$ is the standard Wilson term (with $r=1$).

$|0-\rangle$ is still as in (\ref{eq:0-+}).
Let $V = \pmatrix{ \alpha & \beta \cr \gamma & \delta }$
be the unitary matrix that diagonalizes $\ham_L^+$,
with the first and second ``block-column'' spanning the subspaces of
positive and negative eigenvalues, respectively. 
Then, for a vector theory,
\be
|\<0-|0+\>|^2 = \det \alpha \det \alpha^\dagger .
\ee
This can be obtained as the determinant of the overlap Dirac
operator~\cite{HN1}
\be
D_{\rm ov} = \frac{1}{2} \left[ 1 + \gamma_5 \epsilon(\ham_L^+) \right]
\ee
where $\epsilon(x)$ denotes the sign function. To see this consider,
\be
D_{\rm ov} V = \frac{1}{2} \left[ \pmatrix{ \alpha &
\beta \cr \gamma & \delta } + \gamma_5 \pmatrix{ \alpha &
-\beta \cr \gamma & -\delta } \right] = \pmatrix{ \alpha &
0 \cr 0 & \delta } ~.
\ee
Since $V$ is unitary we have $\det V = \det \delta / \det \alpha^\dagger$
and hence we obtain
\be
\det D_{\rm ov} = \det \alpha \det \alpha^\dagger ~.
\ee

The overlap Dirac operator can be generalized to the massive case
\be
D_{\rm ov}(\mu) = \frac{1}{2} \left[ 1 + \mu + (1-\mu) \gamma_5
 \epsilon(\ham_L^+) \right]
\label{eq:D_ov}
\ee
where $-1 < \mu <1$ is related to the fermion mass by~\cite{EHN1}
\be
m_f = Z_m^{-1} \mu (1 + {\cal O}(a^2)) .
\ee
The propagator for external fermions is given by
\be
{\tilde D}^{-1}(\mu) = (1-\mu)^{-1} \left[ D_{\rm ov}^{-1}(\mu) -1 \right] ,
\label{eq:prop}
\ee
{\it i.e.} it has a contact term subtracted, which makes the massless
propagator chiral: $\{ {\tilde D}^{-1}(0), \gamma_5 \} = 0$.

A massless vector gauge theory can also be obtained from domain wall
fermions~\cite{Kaplan}, where an extra, fifth dimension, of infinite extent
is introduced. 
In the version of ref.~\cite{Shamir},
one can show~\cite{HN3} that the physical (light) fermions contribute
$\log \det D_{\rm DW}$ to the effective action with the 4-d action
\be
D_{\rm DW} = \frac{1}{2} \left[ 1+\mu + (1-\mu) \gamma_5
\tanh \left( - \frac{L_s}{2} \log T \right) \right]
\label{eq:D_DW}
\ee
where $T$ is the transfer matrix in the extra dimension
and $L_s$ its size.
As long as $\log T \ne 0$ we obtain in the limit as $L_s \to \infty$
\be
D_{\rm DW} \rightarrow \frac{1}{2} \left[ 1+\mu + (1-\mu) \gamma_5
 \epsilon( - \log T ) \right].
\ee
This is just the massive overlap Dirac operator up to the replacement
$\ham_w \to - \log T$. It is easy to see that in the limit $a_s \to 0$,
where $a_s$ is the lattice spacing in the extra dimension (set to 1 above),
one obtains $- \log T = \ham_w \left( 1 + {\cal O}(a_s) \right)$.

\section{Some properties of the overlap Dirac operator}

In many cases it is more convenient to use the hermitian version of
the overlap Dirac operator (\ref{eq:D_ov}):
\be
H_o(\mu) = \gamma_5 D_{\rm ov}(\mu) = \frac{1}{2} \left[ (1+\mu) \gamma_5
 + (1-\mu) \epsilon(H_w)  \right] .
\label{eq:H_o}
\ee
The massless version satisfies,
\be
\{H_o(0), \gamma_5 \} = 2 H_o^2(0) .
\ee
It follows that $[H_o^2(0), \gamma_5] = 0$, {\it i.e.} the eigenvectors
of $H_o^2(0)$ can be chosen as chiral. Since
\be
H_o^2(\mu) = ( 1 - \mu^2 ) H_o^2(0) + \mu^2
\ee
this holds also for the massive case.

The only eigenvalues of $H_o(0)$ with chiral eigenvectors are 0 and $\pm 1$.
Each eigenvalue $0 < \lambda^2 < 1$ of $H_o^2(0)$ is then doubly degenerate
with opposite chirality eigenvectors. In this basis $H_o(\mu)$ and
$D_{\rm ov}(\mu)$ are block diagonal with $2 \times 2$ blocks, {\it e.g}
\be
D_{\rm ov}(\mu) : \quad \pmatrix{
    (1-\mu) \lambda^2 + \mu & (1-\mu) \lambda \sqrt{1-\lambda^2} \cr
    -(1-\mu) \lambda \sqrt{1-\lambda^2} & (1-\mu) \lambda^2 + \mu
    } ,
\ee
where
\be
\gamma_5 = \pmatrix{ 1 & 0 \cr 0 & -1 } .
\ee

For a gauge field with topological charge $Q \ne 0$, there are, in addition,
$|Q|$ exact zero modes with chirality ${\rm sign}(Q)$, paired with
eigenvectors of opposite chirality and eigenvalue 1. These are also
eigenvectors of $H_o(\mu)$ and $D_{\rm ov}(\mu)$:
\be
D_{\rm ov}(\mu)_{\rm zero~sector} : \quad \pmatrix{
                         \mu & 0 \cr 0 & 1 }
  \qquad {\rm or} \qquad
  \pmatrix{ 1 & 0 \cr 0 & \mu }
\ee
depending on the sign of $Q$.

We remark that from eigenvalues/vectors of $H_o^2(0)$ those of both
$H_o(\mu)$ and $D_{\rm ov}(\mu)$ are easily obtained. There is no need
for a non-hermitian eigenvalue/vector solver! For example, the Ritz
algorithm~\cite{ritz} will do just fine.

\section{Small eigenvalues and the chiral condensate}

In the chiral eigenbasis of $H_o^2(0)$ the external propagator takes the
block diagonal form with $2 \times 2$ blocks
\be
{\tilde D}^{-1}(\mu) : \quad \frac{1}{\lambda^2 (1-\mu^2) + \mu^2}
         \pmatrix{
         \mu (1-\lambda^2) & -\lambda \sqrt{1-\lambda^2} \cr
         \lambda \sqrt{1-\lambda^2} & \mu (1-\lambda^2)
         } ,
\ee
and, in topologically non-trivial background fields the $|Q|$ additional
blocks, depending on the sign of $Q$,
\be
\pmatrix{ \frac{1}{\mu} & 0 \cr 0 & 0 }
  \qquad {\rm or} \qquad
\pmatrix{ 0 & 0 \cr 0 & \frac{1}{\mu} } .
\ee

We thus find in a fixed gauge field background
\be
\< \bar \psi \psi \>(\{U\}) = \frac{|Q|}{\mu V} 
 + \frac{1}{V} \sum_{\lambda > 0} \frac{2\mu (1-\lambda^2)}
 {\lambda^2 (1-\mu^2) + \mu^2} ,
\ee
and averaged over gauge fields we get the condensate.
It is dominated by the small (non-zero) eigenvalues and in the thermodynamic
limit, where the first term vanishes, it is given by the density of
eigenvalues at zero, $\rho(0^+)$.

With our normalizations we find for all chiral vectors $| b \rangle$
\be
\mu \langle b | \Bigl[ \gamma_5{\tilde D}^{-1}(\mu) \Bigr]^2 | b \rangle
= \langle b | {\tilde D}^{-1}(\mu) | b \rangle
\quad \forall b \quad{\rm with} \quad
\gamma_5 | b \rangle= \pm | b\rangle~.
\ee
This ensures the relation $\mu \chi_\pi = 2 \< \bar \psi \psi \>$ for
every configuration, and, in fact, for every chiral random source
used in a stochastic estimation of condensate and chiral susceptibility
$\chi_\pi$. For such stochastic estimates,
we always work in the chiral sector with no zero-modes.

\section{Implementations of the overlap Dirac operator}

In practice, we only need the application of $D(\mu)$ on a vector,
$D(\mu) \psi$, and therefore only the sign function applied to a
vector, $\epsilon(H_w) \psi$. Since we need the sign function of
an operator (a large sparse matrix) this is still a formidable task.

\begin{figure}
\begin{center}
\epsfysize 50mm
\centerline{\epsfbox[50 20 570 320]{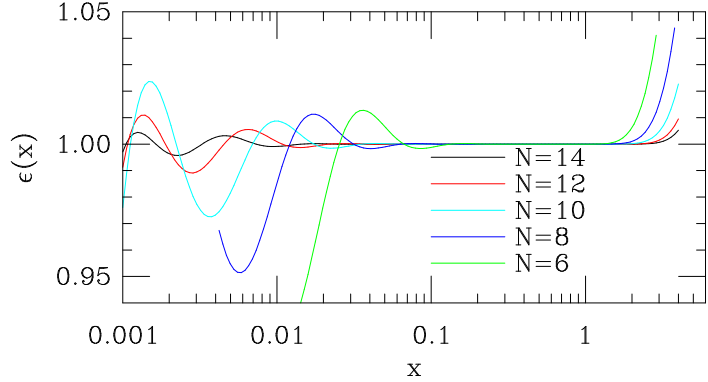}}
\caption{Plots of the optimal rational function approximation to
$\epsilon(x)$ for various order polynomials.}
\label{fig:Remez}
\end{center}
\end{figure}

Methods proposed for this computation are:
\begin{itemize}
\item A Chebyshev approximation of $\epsilon(x)=\frac{x}{\sqrt{x^2}}$
over some interval $[\delta,1]$~\cite{HJL}. For small $\delta$ a large
number of terms are needed.

\item A fractional inverse method using Gegenbauer polynomials for
$\frac{1}{\sqrt{x^2}}$~\cite{Bunk}. This has a poor convergence since
these polynomials are not optimal in the Krylov space.

\item Use a Lanczos based method to compute $\frac{1}{\sqrt{x^2}}$ based
on the sequence generated for the computation of
$\frac{1}{x}$~\cite{Borici}.

\item Use a rational polynomial approximation for $\epsilon(x)$ which
can then be rewritten as a sum over poles:
\be
\epsilon(x) \leftarrow x \frac{P(x^2)}{Q(x^2)} = 
  x \Bigl(c_0 + \sum_k \frac{c_k}{x^2 + b_k}\Bigr)
\ee
The application of $\chi\leftarrow\epsilon(H_w)\psi$ can be done by
the simultaneous solution of the shifted linear systems~\cite{Jegerlehner}
\be
(H_w^2 + b_k)\phi_k = \psi,\qquad \chi = H_w (c_0 \psi + \sum_k c_k \phi_k) .
\ee
One such approximation, based on the polar decomposition~\cite{pensacola},
was introduced in this context by Neuberger~\cite{HN2}.
We use optimal rational polynomials~\cite{EHN2}. The accuracy of this
approximation is shown in Fig.~\ref{fig:Remez}.
\end{itemize}

We note that in all methods listed above, one can enforce the accuracy
of the approximation of $\epsilon(x)$ for small $x$ by projecting out
the lowest few
eigenvectors of $H_w$ and adding their correct contribution exactly. 
\be
\epsilon(H_w) = \sum_{i=1}^n |\psi_i\> \epsilon(\lambda_i) \<\psi_i|
 + {\cal P}_\perp^{(n)} {\rm App}[\epsilon(H_w)] {\cal P}_\perp^{(n)} ,
\qquad {\cal P}_\perp^{(n)} = {\bf 1} - \sum_{i=1}^n |\psi_i\> \<\psi_i| .
\label{eq:project}
\ee

To invert $D^\dagger D$ for overlap fermions, we have, generically, an outer
CG method (a 4-d Krylov space search) and an independent inner search method
for $\epsilon(H_w) \psi$ -- maybe CG again. For domain wall fermions, on the
other hand, a 5-d Krylov space search method is used.
It may pay off to try to combine inner and outer CGs for overlap
fermions by reformulating them into a 5-d problem~\cite{HN4,Borici2}.

\section{Main problem for Overlap and Domain Wall fermions}

\begin{figure}
\begin{center}
\epsfysize 65mm
\centerline{\epsfbox[0 0 630 580]{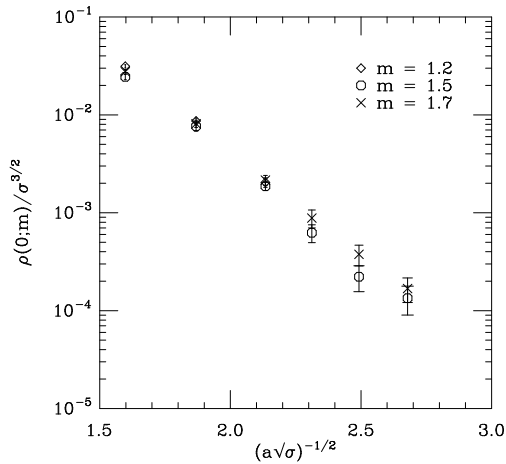}}
\caption{The approach of $\rho(0;m)$ to the continuum limit in the
quenched theory.}
\label{fig:rho_0}
\end{center}
\end{figure}

For topology to change, we must create dislocations. These produce small
modes which force the spectral gap of $H_w(m)$ to be closed. The density
of zero eigenvalues of $H_w(m)$, $\rho(0;m)$, is non-zero in the quenched
case, but rapidly decreasing with decreasing coupling~\cite{EHN3}.
Very roughly, we find $\rho(0;m)/\sigma^{3/2} \ \sim \ e^{-e^\beta}$ as
shown in Fig.~\ref{fig:rho_0}.

The existence of small eigenvalues hampers the approximation accuracy and
convergence properties of implementations of $\epsilon(H_w)$. Eigenvector
projection both increases the accuracy of the approximation and
decreases the condition number, {\em e.g.} of the inner CG.

The existence of small eigenvalues has implications also for domain
wall fermions. One can show that the spectrum of $-\log{T(m)}$ of
Eq.~(\ref{eq:D_DW}) around zero is the same as the spectrum of
$H_w(m)$~\cite{overlap}. While the small eigenvalues of $-\log{T(m)}$
don't appear to cause algorithmic problems for domain wall fermions,
they can induce rather strong $L_s$ dependence of physical quantities,
and causing hence the need for large $L_s$.

\section{The Dirac spectrum, chiral condensate and chiral Random Matrix Theory}

Up to a scale, given by the infinite volume chiral condensate
$\Sigma=\<\bar\psi\psi\>$, RMT predicts that the rescaled density of
eigenvalues
\be
\rho_S(z) = \lim_{V \to \infty} \frac{1}{V} \rho \left(
 \frac{z}{V\Sigma} \right)
\ee
is universal, dependent only on the symmetry properties, number of
dynamical flavors, and the number of exact zero modes (the topological
sector), but not the form of the potential in the random matrix
theory, or low energy effective Lagrangian~\cite{RMT_review}.
There are three classes of random matrices determined by their symmetry
properties: orthogonal, unitary, and symplectic.

\begin{figure}
\begin{center}
\epsfysize 42mm
\centerline{\epsfbox[70 230 590 405]{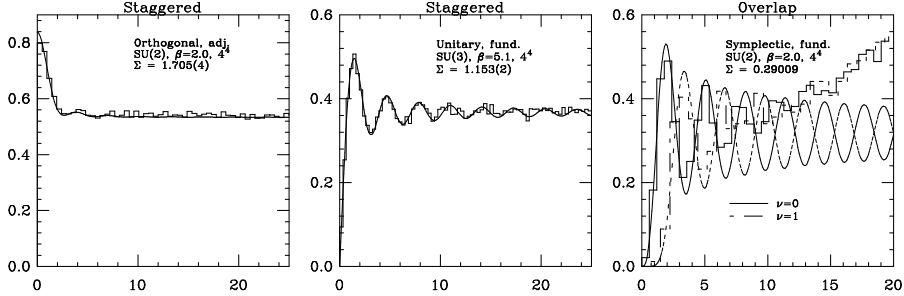}}
\caption{Examples for the microscopic spectral density for all three
ensembles. For overlap fermions one can probe different topological
sectors (rightmost panel).}
\label{fig:rho_S}
\end{center}
\end{figure}

In Fig.~\ref{fig:rho_S} we show examples of the microscopic spectral
density for all three ensembles and compare to the analytic predictions
from RMT. With overlap fermions we can probe topologically non-trivial
sectors.

\begin{figure}
\begin{center}
\epsfysize 70mm
\centerline{\epsfbox[80 200 550 500]{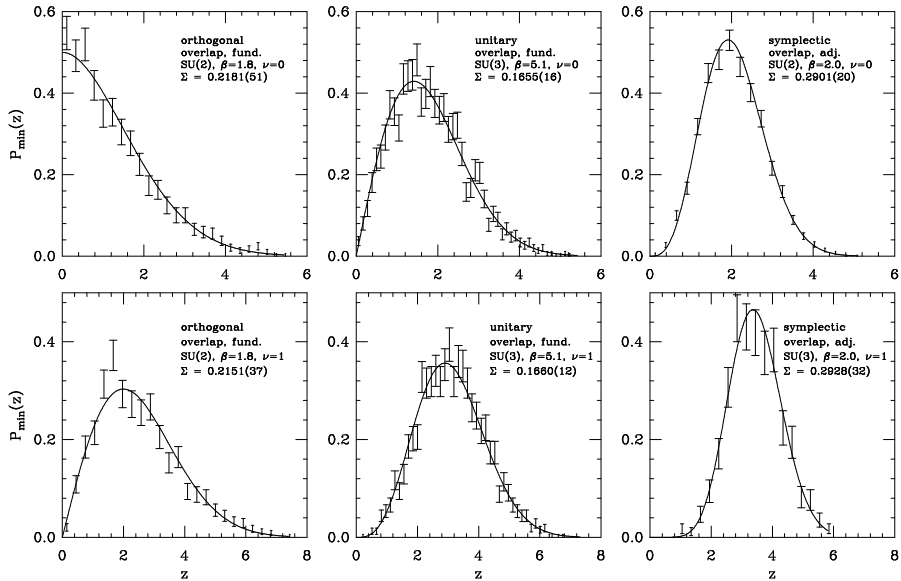}}
\vspace*{10mm}
\caption{Plots of the distribution of the lowest eigenvalue for all three
ensemble in the lowest two topological sectors. The curves are fits to
the predictions from random matrix theory.}
\label{fig:pmin}
\end{center}
\end{figure}

Similarly, there are predictions in each ensemble and topological sector
for the distribution of the lowest eigenvalue.
Examples for the quenched theory with overlap fermions are shown in
Fig.~\ref{fig:pmin}. The $\Sigma$'s extracted from fits in different
$\nu$ sectors for each ensemble are consistent~\cite{EHKN1}.

\begin{figure}
\begin{center}
\epsfysize 42mm
\centerline{\epsfbox[50 230 590 405]{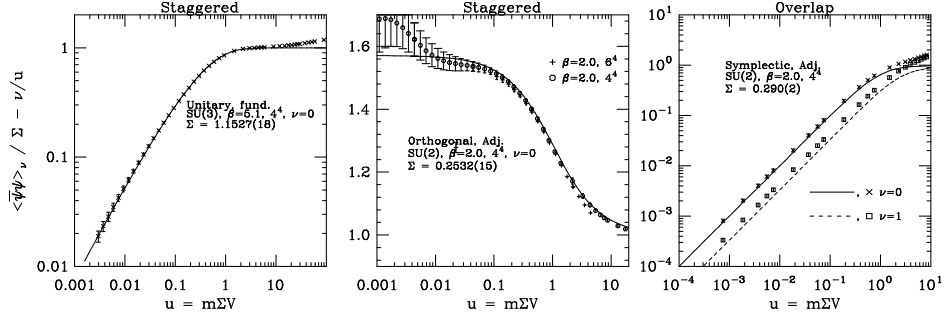}}
\caption{Examples of the behavior of the chiral condensate and the
comparison to predictions from random matrix theory.}
\label{fig:pbp_rmt}
\end{center}
\end{figure}

RMT also gives predictions for the finite mass and volume dependence of the
chiral condensate in the small mass large volume regime,
\be
\frac{\Sigma_\nu(u)}{\Sigma} = 2 u \! \int_0^\infty dz
\frac{\rho_S(z)}{z^2+u^2}\quad +\frac{\nu}{u} ~,
\ee
with $u = \mu \Sigma V$.
Particularly interesting is the behavior at small $u$:
\be
\Sigma_0^{\rm GUE}(u)/\Sigma \sim -u\log{u},\quad
\Sigma_0^{\rm GOE}(u)/\Sigma \sim \frac{1}{2}(\pi - u),\quad
\Sigma_{0,1}^{\rm GSE}(u)/\Sigma \sim u
\ee
It is very sensitive to the lowest eigenvalues.
In quenched QCD, surprisingly, $\Sigma_{0}^{\rm GOE}(u)/\Sigma$ does not
vanish at $u=0$ in this microscopic limit.
Our data, shown in Fig.~\ref{fig:pbp_rmt} follow the predictions
well~\cite{DEHN}. Once again, with
overlap fermions we can probe topologically non-trivial sectors.
The finite volume corrections are quite large for overlap fermions: $\Sigma$
is about a factor $7$ smaller than in the staggered case. This implies
that for overlap fermions larger volumes are needed to see the microscopic
regime.

\section{ Small eigenvalue distribution in quenched QCD above $T_c$}

\begin{figure}
\begin{center}
\epsfxsize 100mm
\centerline{\epsfbox[10 60 640 570]{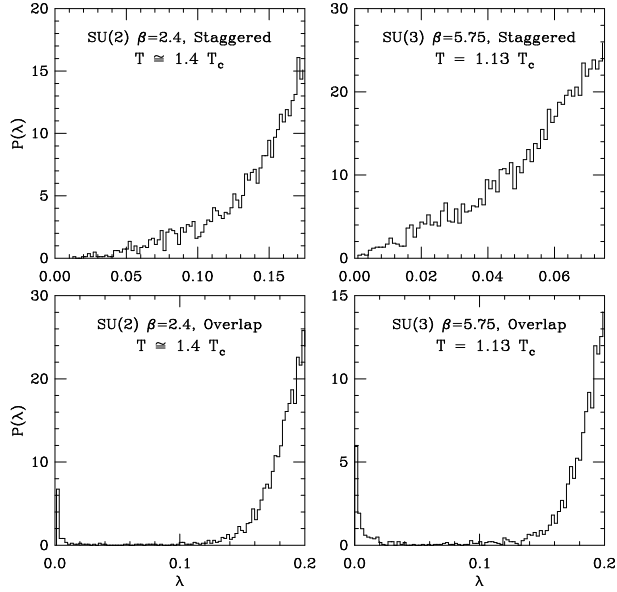}}
\caption{Low lying eigenvalue distributions in quenched QCD at finite
temperature for staggered and overlap fermions.}
\label{fig:ft_evs}
\end{center}
\end{figure}

We have studied the small eigenvalue distribution of the Dirac operator
in the deconfined phase of quenched QCD.
Sample distributions of small (non-zero) eigenvalues are shown in
Fig.~\ref{fig:ft_evs}~\cite{EHKN2}.

For overlap fermions, we see the lower end of the bulk of the distribution,
then a dip, or even a gap, and then again small eigenvalues, below about 0.05.
We focus on the small modes, $\lambda < 0.05$. Our findings are summarized
in Tables~\ref{tab:su3} and \ref{tab:su2}.
We see that both $\<n\>/V$ and $\<Q^2\>/V$ seem to remain finite and non-zero
in the large volume limit for fixed $\beta$, but they drop quickly as
$\beta$, and hence the temperature, is increased.

\begin{table}[htb]
\begin{center}
\caption{
SU(3) data: $n=n_++n_-$ with $n_\pm$ the number of zero and small non-zero
eigenvalues with chirality $\pm$.
$Q=n_+-n_-$ is the topological charge. $\sigma_n$ is the variance of $n$. 
The volume normalizations for $n$ and $Q^2$ are per spatial $8^3$ volume.
}
\label{tab:su3}
\begin{tabular}{c|cc|c|ccc} \hline
 volume           &\multicolumn{2}{|c|}{$8^3\times 4$} & $12^3\times 4$
                  &\multicolumn{3}{|c}{$16^3\times 4$} \\
 $\beta$          & 5.75 & 5.85 & 5.75 & 5.71 & 5.75 & 5.85 \\ \hline
 $\<n\>/V$        & 0.32 & 0.06 & 0.28 & 0.63 & 0.30 & 0.05 \\
 $\<Q^2\>/V$      & 0.31 & 0.07 & 0.28 & 0.64 & 0.33 & 0.05 \\
 $\<n\>/\sigma_n$ & 1.09 & 0.90 & 0.92 & 1.15 & 1.03 & 0.83 \\ \hline
\end{tabular}
\end{center}
\end{table}

\begin{table}[htb]
\begin{center}
\caption{SU(2) data}
\label{tab:su2}
\begin{tabular}{c|cc|cc} \hline
 volume           &\multicolumn{2}{|c|}{$8^3\times 4$}
                  &\multicolumn{2}{|c}{$16^3\times 4$} \\
 $\beta$          & 2.3  & 2.4  & 2.4  & 2.5  \\ \hline
 $\<n\>/V$        & 1.66 & 0.29 & 0.25 & 0.05 \\ 
 $\<Q^2\>/V$      & 1.66 & 0.31 & 0.25 & 0.05 \\
 $\<n\>/\sigma_n$ & 0.97 & 1.09 & 0.93 & 0.99 \\ \hline
\end{tabular}
\end{center}
\end{table}

Looking in more detailed at the small modes we find
\begin{itemize}
  \item Their number $n$ is roughly Poisson distributed, $ P(n,\<n\>) = 
        \<n\>^n e^{-\<n\>}/n!$.

        Average and variance are approximately equal.
  \item For fixed $n$, $n_+$ and $n_-$ are roughly binomially distributed.
\end{itemize}
These observations are consistent with interpreting the small modes to
be due to a dilute gas of instantons and anti-instantons, with $n_+$ and
$n_-$ their numbers. $n-|Q|$ of the would-be zero modes mix due to their
overlapping and get small eigenvalues, while $|Q|$ exact zero modes remain.

At finite temperature, instantons fall off exponentially, and so do the
fermionic zero modes associated with them. We consider a toy model of
randomly (Poisson and binomially) distributed instantons and
anti-instantons, inducing interactions of the form
$h_0 {\rm e}^{-d(i,j)/D}$ between the would-be zero modes of every instanton
-- anti-instanton pair $(i,j)$ with separation $d(i,j)$. Like sign pairs
are assumed to have no interactions. This toy model reproduces all
qualitative features of the small eigenvalue distributions well for $D
\approx 2$, corresponding to $D \approx 1/(2T)$~\cite{EHKN2}.

\section{ Conclusions}

The overlap Dirac operator has the same chiral symmetries as
continuum fermions. It has exact zero modes in topologically
non-trivial gauge fields. It is therefore well suited for a study
of the interplay of topology, with its
associated exact zero modes, and chiral symmetry breaking,
determined by the density of small eigenvalues.

In the range of its validity the predictions of chiral random
matrix theory are well followed and confirmed by overlap
fermions, including the dependence on topology, given by the number
of exact zero modes.

A study of the small eigenvalues in quenched QCD above the
deconfining transition temperature, $T_c$, shows that topology,
manifested by exact zero modes, persists.
Furthermore, a finite density of small eigenvalues persists,
and their properties are well described by attributing them
to the would-be zero modes of a random dilute gas of instantons
and anti-instantons.

\section*{Acknowledgements}
The work of RGE and UMH has been supported in part by DOE contracts
DE-FG05-85ER250000 and DE-FG05-96ER40979.

\end{document}